\documentclass[11pt,a4paper]{article}
\usepackage{jcappub}

\usepackage{url}
\usepackage{lscape}

\def\beq{\begin{equation}}
\def\eeq{\end{equation}}
\def\alwaysmath#1{{\ifmmode{#1}\else{$#1$}\fi}}
\def\he#1{\hbox{\alwaysmath{{}^{#1}}{\rm He}}}

\title{
The primordial helium abundance from updated emissivities
}

\author[a]{Erik Aver}
\author[b,c,d]{Keith~A.~Olive}
\author[e,f]{R.L.~Porter}
\author[b,d]{Evan~D.~Skillman}

\affiliation[a]{Department of Physics, Gonzaga University, \\
502 E Boone Ave, Spokane, WA 99258}
\emailAdd{aver@gonzaga.edu}
\affiliation[b]{School of Physics and Astronomy, University of Minnesota, \\
116 Church St. SE, Minneapolis, MN 55455}
\affiliation[c]{William I. Fine Theoretical Physics Institute, University of Minnesota, \\
116 Church St. SE, Minneapolis, MN 55455}
\emailAdd{olive@umn.edu}
\affiliation[d]{Minnesota Institute for Astrophysics, University of Minnesota, \\
116 Church St. SE, Minneapolis, MN 55455}
\emailAdd{ryanlporter@gmail.com}
\affiliation[e]{Department of Physics and Astronomy, University of Georgia, \\
Athens, GA 30602}
\emailAdd{skillman@astro.umn.edu}
\affiliation[f]{Center for Simulational Physics, University of Georgia, \\
Athens, GA 30602}

\abstract{
Observations of metal-poor extragalactic H~II regions allow the determination of the primordial helium abundance, Y$_{p}$.  The He~I emissivities are the foundation of the model of the H~II region's emission.  Porter, Ferland, Storey, \& Detisch (2012) have recently published updated He~I emissivities based on improved photoionization cross-sections.  We incorporate these new atomic data and update our recent Markov Chain Monte Carlo analysis of the dataset published by Izotov, Thuan, \& Stasi\'nska (2007).  As before, cuts are made to promote quality and reliability, and only solutions which fit the data within 95\% confidence level are used to determine the primordial He abundance.  The previously qualifying dataset is almost entirely retained and with strong concordance between the physical parameters.  Overall, an upward bias from the new emissivities leads to a decrease in Y$_{p}$. In addition, we find a general trend to larger uncertainties in individual objects (due to changes in the emissivities) and an increased variance (due to additional objects included).
From a regression to zero metallicity, we determine Y$_{p}$ $=$ 0.2465 $\pm$ 0.0097, in good agreement with the Planck result of Y$_{p}$ $=$ 0.2485 $\pm$ 0.0002.  
In the future, a better understanding of why a large fraction of spectra are not well fit by the model will be crucial to achieving an increase in the precision of the primordial helium abundance determination.
}


\arxivnumber{}

\begin{document}

\begin{flushright}UMN-TH-3302/13\\FTPI-MINN-13/31\\
August 2013\end{flushright}

\maketitle
\flushbottom

\section{Introduction}

Planck's determination of the cosmological parameters has brought precision cosmology to a new level \citep{planck}.  In particular their determination of the baryon density of $\Omega h^2 = 0.02218 \pm 0.00026$ corresponding to a baryon-to-photon ratio of $\eta = (6.08 \pm 0.07) \times 10^{-10}$.  As a result, standard big bang nucleosynthesis (SBBN)  has become a parameter free theory \citep{cfo2}. Nevertheless, SBBN remains the most precise robust probe of the very early universe available \citep{wssok,osw,fs}.  Using the density as determined by Planck \citep{planck}, one can now
make relatively precise predictions of the initial abundances of the light elements D, $^{3}$He, $^{4}$He, and $^{7}$Li \citep{cfo,coc,cfo3,coc2,cyburt,coc3,cuoco,serp,cfo5,coc4,cuv}.  Any test of SBBN, therefore, requires abundance determinations at suitably high precision.  At the Planck determined value of $\eta$, and a neutron mean life of $880.1 \pm 1.1$ s \citep{rpp}, SBBN yields $Y_p = 0.2485 \pm 0.0002$, a relative uncertainty of only 0.08\% \citep{cyburt,cfo5}.

Prior to the discovery of deuterium in quasar absorption systems \citep{deut}, helium abundance determinations in low metallicity H~II regions in dwarf galaxies provided the most stringent test of SBBN and still provides important constraints on the physics of the early universe beyond the standard model \citep{cfos}. While the \he4 abundance is certainly the most accurately determined of the light element abundances (though significant advances have been made for D/H \citep{cooke}), obtaining better than 1\% accuracy on the determination in individual objects remains a challenge.  The primordial abundance of \he4, Y$_{p}$, is determined by fitting the helium abundance versus metallicity, and extrapolating back to very low metallicity \citep{ptp74}.  The oxygen to hydrogen ratio, O/H, commonly serves as a proxy for metallicity.  Due to numerous systematic uncertainties, difficulties in extracting an accurate and precise measure of the primordial helium abundance are well established \citep{os01,os04,its07,plp07}. 

The first step in achieving accurate \he4 abundances made use of a ``self-consistent'' analysis method \citep{itl94,itl97}, which fit simultaneously the \he4 abundance from five emission lines along with two key physical parameters associated with the H~II region: electron density, $n_e$ and  optical depth $\tau$. It was argued that the temperature, $T$, could also be solved for in a self-consistent manner \citep{ppr00}.  The importance of Monte Carlo techniques was demonstrated  using six helium and four hydrogen lines \citep{os01,os04}. In addition to $n_e$, $\tau$, and $T$, physical parameters associated with underlying stellar H and He absorption, $a_H$ and $a_{He}$, and reddening, $C(H\beta$) were also determined; however, parameters associated with He and H were solved for separately.  Combined Monte Carlo solutions for H and He were found in \citet[][AOS]{AOS}, where the fraction of neutral hydrogen, $\xi$, was also added to the list of physical parameters.  The current statistical technique using a Markov Chain Monte Carlo (MCMC) analysis to explore a global likelihood function for all model parameters including the helium abundance was established in \citet[][AOS2 \& AOS3, respectively]{AOS2, AOS3}. The full list of model parameters is therefore eight being determined by nine emission line ratios (all with respect to H$\beta$).

Integral to the determination of the \he4 abundance is an accurate set of emissivities used to convert the raw observed flux measurements into a calculable abundance.  Considerable progress has been made over the past 10-15 years in calculated emission line intensities.  Beginning with the Smits overhaul \citep{smits91,smits96} of the long-used Brocklehurst emissivities \citep{brock}, there have been several stages of improvements in theoretical treatment of the He emissivities. The Smits recombination model was combined with the collisional transition rates of Sawey \& Berrington \citep{sb93} in \citet{bss99}. The next major improvement came in \citet{por05} which used improved radiative and collisional data.  These were applied to abundance determinations in \citet[][PFM]{pfm07}.  AOS, AOS2, \& AOS3 utilized the emissivities of PFM.  Recently, Porter, Ferland, Storey, \& Detisch \citep[][PFSD]{pfsd,pfsdc} have published updated emissivities reflecting improved photoionization cross-sections. 

Here, we incorporate the new PFSD emissivities to recalculate the helium abundance in extragalactic H~II regions. As in AOS3, we begin with the large HeBCD dataset of \citet[][ITS07]{its07}.  As explained in AOS3, we consider only those objects with measured $\lambda4026$ emission lines.  For each object, we fit the eight physical parameters (including the \he4 abundance) using the MCMC method and determine the $\chi^{2}$ of the fit. We then select those objects with $\chi^{2}<4$. This set of 16 (an increase from 14 in AOS3) objects is used to extrapolate to zero metallicity and determine the primordial \he4 abundance. 
As we will see, the new emissivities result in a systematic shift to slightly lower abundances with slightly larger uncertainties, with a corresponding effect on the determined primordial \he4 abundance.

This paper is organized as follows.
First, section \ref{Recap} briefly summarizes the analysis and screening of AOS3.  Second, in \S \ref{Emissivities}, the new PFSD emissivities are introduced and compared to the previous PFM emissivities.  Third, \S \ref {Sample} provides an overview of the sample and the effect of cuts on the dataset for quality and reliability.  Subsequently, Y$_p$ is determined in \S \ref{Results}.  Finally, \S \ref{Conclusion} offers a discussion of the results, their exploration, and of further improvements in the determination of the primordial helium abundance.

\section{Revisiting AOS3} \label{Recap}

The Markov Chain Monte Carlo scans our 8-dimensional parameter space 
mapping out the likelihood function based on the $\chi^2$ given by
\beq
\chi^2 = \sum_{\lambda} {(\frac{F(\lambda)}{F(H\beta)} - {\frac{F(\lambda)}{F(H\beta)}}_{meas})^2 \over \sigma(\lambda)^2},
\label{eq:X2}
\eeq
where the emission line fluxes, $F(\lambda)$, are measured and calculated for 
six helium lines  ($\lambda\lambda$3889, 4026, 4471, 5876, 6678, and 7065) and three
hydrogen lines  (H$\alpha$, H$\gamma$, H$\delta$) each relative to H$\beta$.
The  $\chi^{2}$ in eq. \ref{eq:X2} runs over all He and H lines and $\sigma(\lambda)$ is the
measured uncertainty in the flux ratio at each wavelength. Once minimized, best fit solutions
for the eight physical parameter inputs are found, and uncertainties in each quantity 
can be obtained by calculating a 1D marginalized likelihood. 
In AOS3, He line flux ratios (compared to H$\beta$) were calculated using 
\beq
\frac{F(\lambda)}{F(H\beta)}= y^{+}\frac{E(\lambda)}{E(H\beta)}{\frac{W(H\beta)+a_{H}(H\beta)}{W(H\beta)} \over \frac{W(\lambda)+a_{He}(\lambda)}{W(\lambda)}}{f_{\tau}(\lambda)}\frac{1+\frac{C}{R}(\lambda)}{1+\frac{C}{R}(H\beta)}10^{-f(\lambda)C(H\beta)} ,
\label{eq:F_He_EW}
\eeq
along with an analogous expression for H line flux ratios. 
In eq. (\ref{eq:F_He_EW}), $y^+$ corresponds to the input abundance by number (relative to H)
of ionized He. $W(\lambda)$ is the measured equivalent width and two parameters, $a_H$ and $a_{He}$, characterize the wavelength-dependent underlying absorption for H and He respectively. The function $f_\tau(\lambda)$ represents a correction for florescence. 
In AOS3, a fit for $f_\tau$ was used that includes collisional corrections and depends on $\tau, n_e$, and $T$ \citep{bss02}.  The  emissivity, $E$, and He collisional corrections were taken from PFM \citep{pfm07}. The final term in  eq. (\ref{eq:F_He_EW}), accounts for reddening.

AOS3 analyzed the 93 H~II region observations reported in the HeBCD sample of \citep[][ITS07]{its07}.  Extensive screening was conducted to promote reliability and achieve a robust dataset for determining the primordial helium abundance (please see AOS3 for more detail \citep{AOS3}).  First, observations for which He~I $\lambda$4026 was not detected were excluded to reduce systematic uncertainty due to the underlying helium absorption that may be introduced by the absence of He~I $\lambda$4026.  This left 70 objects in the database. Second, best-fit solutions with $\chi^{2}$ values greater than 4, corresponding to a standard 95\% confidence level, were excluded. This was another large cut, leaving only 25 objects remaining. Third, solutions with unphysical physical parameters, namely $\xi > 0.333$ ($>$ 25\% neutral hydrogen), were excluded (2 more objects).  Finally, to reduce systematic uncertainty due to the assumed linear metallicity relationship between He/H and O/H, objects with $O/H \ge 15.2 \times 10^{-5}$ were also excluded (one additional object excluded).   The $\chi^{2}<4$ criterion itself proved effective at identifying unphysical or ambiguous solutions.  However, it also excluded nearly two thirds of the observations with He~I $\lambda$4026 detected, raising questions into potential deficiencies of the model or data.  Cumulatively, the cuts just specified yielded a dataset with 22 objects.

The 22 objects for which the model was a good fit were also examined and flagged for parameter outliers.  The models for optical depth and underlying absorption carry significant systematic uncertainties.  To limit the effect of these systematic uncertainties, objects with large corrections for these factors were flagged: $\tau > 4$, a$_H > 6$ \AA, a$_{He} > 1$ \AA, and finally, $\xi > 0.01$, where the 1-$\sigma$ lower bound does not encompass $\xi=0.001$.  Furthermore, the solution for the electron temperature should be in relatively good agreement with the temperature derived from the [O~III] emission lines (which is used as a very conservative prior; see AOS2 for further discussion \citep{AOS2}), with T(O~III) serving as a loose upper bound on T.  Thus, screening for objects with $T(O~III) - T > 5000~K$ or $T(O~III) - T < -3000~K$ was also conducted, but none were found.  Of the 22 retained objects, a total of 8 were flagged.  Table \ref{table:Cuts_AOS3} summarizes the cuts and their effects on the dataset in AOS3.  In section \ref{Sample} we redo the analysis of AOS3 with the new PFSD emissivities. We start with the same HeBCD dataset of 70 objects (those with He~I $\lambda$4026 detected), and preform the $\chi^2$ analysis and make the same set of cuts. Those results are all 
shown in table \ref{table:Cuts_AOS3}. 

The net result from AOS3 for the primordial \he4 mass fraction was 
\beq
Y = 0.2534 \pm 0.0083 + (54 \pm 102) {\rm O/H},
\eeq
based on a linear regression of the 14 surviving objects, and
\beq
Y_p = 0.2574 \pm 0.0036,
\eeq
based on a weighted mean of the same data.

\begin{table}[ht!]
\centering
\vskip .1in
\begin{tabular}{lrr}
\hline\hline
HeBCD (ITS07)				& 93  & 93 \\
\hline
Emissivities & PFM & PFSD \\
\hline
\underline{\textit{Cuts}} \\
He~I $\lambda$4026 Not Detected		& 23/93 & 23/93 \\
$\chi^{2}>4$		   		& 45/70 & 35/70 \\
$\xi>0.333$				&  2/25 &  7/35	\\
Degenerate Solution			&  0/23 &  0/28	\\
$O/H \ge 15.2 \times 10^{-5}$		&  1/23 &  1/28	\\
\hline
Subtotal:  Well-Defined Solutions	&    22 &    27 \\
\hline
\underline{\textit{Flagged}} \\
\hspace{0.1in} $\xi>0.01$ \& $(\xi-\sigma(\xi))>0.001$	&  3 & 7 \\
\hspace{0.1in} $\tau>4$					&  1 & 2 \\
\hspace{0.1in} a$_H > 6$ \AA				&  3 & 3 \\
\hspace{0.1in} a$_{He} > 1$ \AA				&  2 & 3 \\
\hspace{0.1in} $T(O~III) - T < -3000~K$			&  0 & 0 \\
\hspace{0.1in} $T(O~III) - T >  5000~K$			&  0 & 0 \\
\hline
Subtotal:  Flagged					&  8 & 11 \\
\hline
Final Dataset				& 22  & 27 \\
\hspace{0.1in} Flagged			&  8  & 11 \\
\hspace{0.1in} Qualifying		& 14  & 16 \\
\hline
\end{tabular}
\caption{Breakdown of the cuts and flags on the HeBCD dataset analyzed in AOS3 and here, employing
either the PFM or PFSD set of emissivities, respectively.}
\label{table:Cuts_AOS3}
\end{table}

\section{Investigating the updated emissivities} \label{Emissivities}

Both sets of emissivities, PFM \& PFSD (\citep{pfm07,pfsd}), were calculated for the Case B approximation with the spectral simulation code CLOUDY,
last described by \citet{fer13}.  The main sources of uncertainty in these calculations are the photoionization cross-sections and the collisional rates \citep{bss99, por05, pfm07}.  For primordial helium analysis, photoionization cross-sections may contribute the dominant source of uncertainty \citep{pfms09}.  

\begin{figure}
\centering  
\resizebox{\textwidth}{!}{\includegraphics{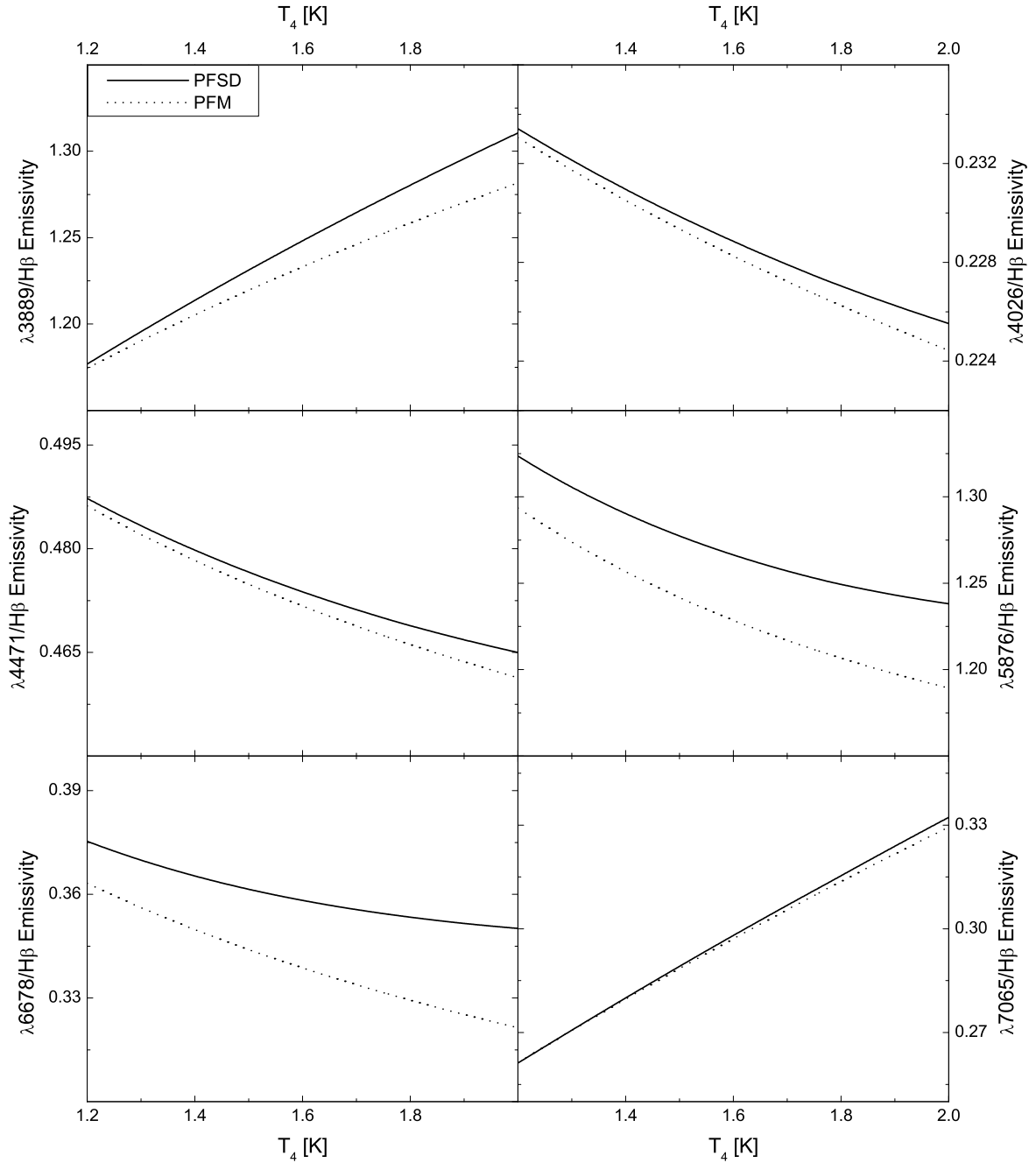}}
\caption{
Comparison of the PFSD emissivities, $\frac{{\tilde E}(\lambda)}{E(H\beta)}$, to those of PFM, including the collisional correction, $\frac{E(\lambda)}{E(H\beta)}(1+\frac{C}{R}(\lambda))$ (both for n$_e$ = 100 cm$^{-3}$).  The PFM fits are the dashed lines while the PFSD fits are solid.  The progression is, left to right, top to bottom, by wavelength: $\lambda\lambda$3889, 4026, 4471, 5876, 6678, 7065.
}
\label{Emissivity}
\end{figure}

\begin{figure}
\centering  
\resizebox{\textwidth}{!}{\includegraphics{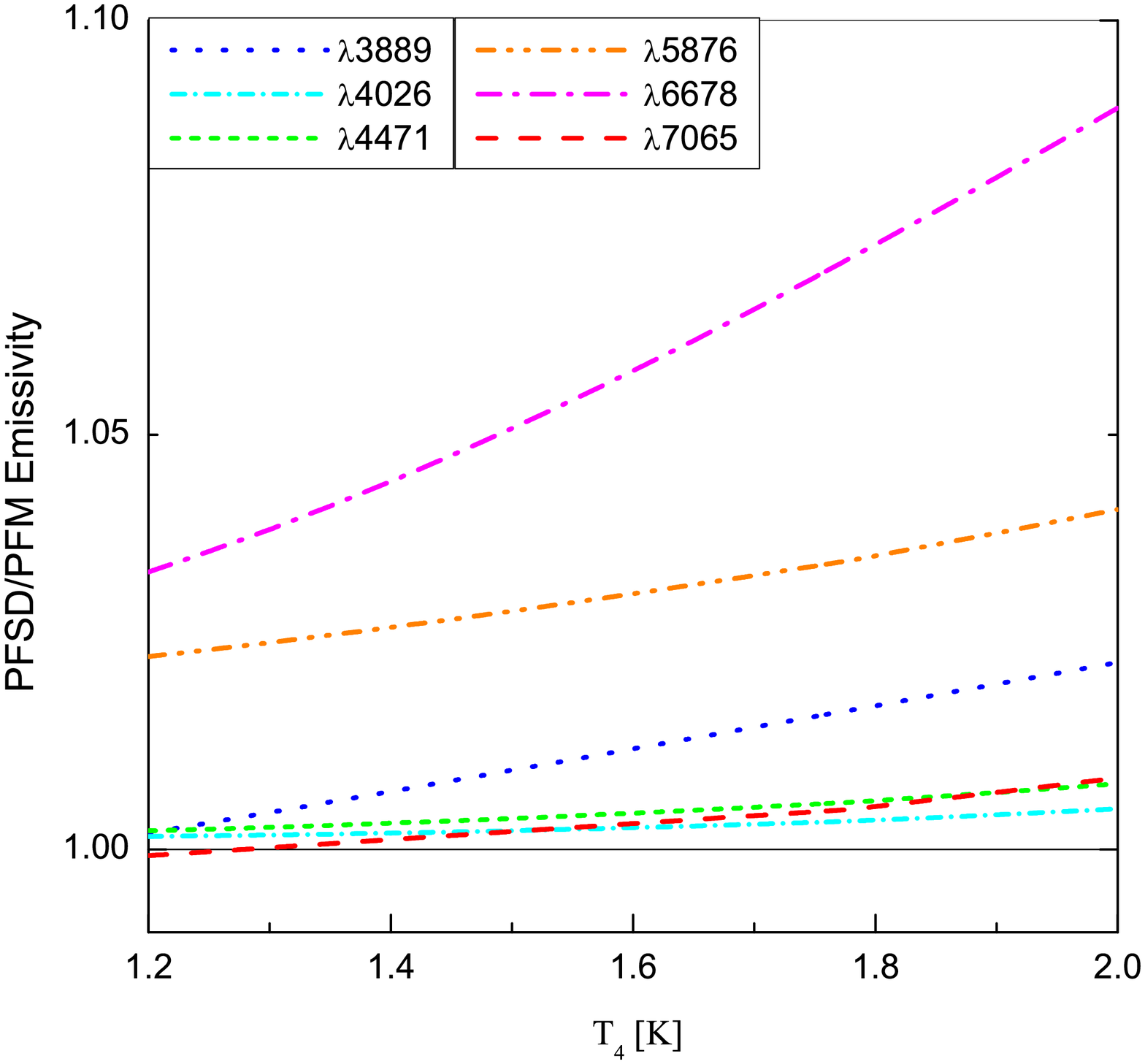}}
\caption{
The PFSD emissivities, $\frac{{\tilde E}(\lambda)}{E(H\beta)}$, plotted relative to those of PFM, including the collisional correction, $\frac{E(\lambda)}{E(H\beta)}(1+\frac{C}{R}(\lambda))$ (both for n$_e$ = 100 cm$^{-3}$).  The updated emissivities all show an increase compared to the older emissivities, but the relative shifts are clearly not the same for all six lines.  He~I $\lambda\lambda$4026, 4471, and 7065, all show similar, small increases of $<1\%$, but $\lambda\lambda$5876 and 6678 show significantly larger increases.
}
\label{EmissivityRelative}
\end{figure}

The new emissivities presented in PFSD differ from the earlier PFM results in two main ways.  
First, photoionization cross-sections were improved with a larger ab initio set from \citet{hs98}.  
While this advance was important in its own right, it also revealed a programming error in the earlier treatment.  
Correcting the error resulted in $\sim 3\%$ increases in the effective recombination coefficients into 
the $3d^{1}D$ and $3d^{3}D$ levels (the upper levels of $\lambda\lambda$5876 and 6678).  This is the dominant effect in figures \ref{Emissivity} and \ref{EmissivityRelative}, which show a comparison of PFSD and PFM emissivities.

Figure \ref{Emissivity} provides a comparison of the PFSD and PFM emissivities versus temperature for all six helium lines.  Figure \ref{EmissivityRelative} is similar, except that it overplots the PFSD emissivities relative to the PFM emissivities for all six lines.  As both figures illustrate, the PFSD emissivities increased for all six lines.  While the He~I $\lambda\lambda$4026, 4471, and 7065 emissivities all increase by $<1\%$,  $\lambda\lambda$5876 and 6678 increase more significantly, exceeding $5\%$ for $\lambda$6678 at higher temperatures.  Note also that the temperature dependencies of the new $\lambda\lambda$5876 and 6678 emissivities lessen at higher temperatures compared to the older emissivities.  This results in less stringent constraints on the derived temperature, and thus, a larger uncertainty in the derived helium abundances.  

The second effect is reported for the first time here, as it was forgotten for several years and rediscovered in the course of preparing this manuscript.  
The CLOUDY treatment of electron-induced collisions (above some low-lying ab initio calculations) 
is an implementation of empirical results from \citet[][VS80]{vs80}.  
However, in 2008, this implementation was switched from integrating an equation for cross-sections (VS80, eqs. 14 \& 21) 
to direct evaluation of an equation for rate coefficients (VS80, eqs. 17 \& 24).  
The latter form is preferred by Vriens \& Smeets and is computationally faster.  
This change is responsible for the gradual divergence of PFSD and PFM emissivities in figures \ref{Emissivity} and \ref{EmissivityRelative}.

For the purposes of this paper, and in order to minimize interpolation errors, we have calculated He~I emissivities on a much finer parametric mesh than published in PFSD.  Temperatures are in 250~K increments from 10,000~K to 25,000~K (inclusive), and 31 electron densities are irregularly spaced from 1 to 10,000~cm$^{-3}$ (also inclusive).  This finer grid is available upon request from the authors.

The replacement of the PFM He~I emissivities and collisional corrections with the updated results of PFSD is the only change in the present analysis from AOS3.  The new PFSD emissivity data incorporates the collisional contribution to the recombination rates.  As a result, a separate term correcting for the collisional contribution is not required, and eq. (\ref{eq:F_He_EW}) for the ratio of the He flux relative to H$\beta$ is now given as
\beq
\frac{F(\lambda)}{F(H\beta)}= y^{+}\frac{{\tilde E}(\lambda)}{E(H\beta)}{\frac{W(H\beta)+a_{H}(H\beta)}{W(H\beta)} \over \frac{W(\lambda)+a_{He}(\lambda)}{W(\lambda)}}{f_{\tau}(\lambda)}\frac{1}{1+\frac{C}{R}(H\beta)}10^{-f(\lambda)C(H\beta)},
\label{eq:F_He_PFSD}
\eeq
where $\frac{{\tilde E}(\lambda)}{E(H\beta)}$ is the combined emissivity and collisional correction for He and is a function of temperature and density.  
As discussed above, the PFSD emissivities increased for all six lines relative to PFM (see figures \ref{Emissivity} \& \ref{EmissivityRelative}). Therefore, as eq. \ref{eq:F_He_PFSD} shows, to achieve the same flux, an increase in the He~I emissivities tends to decrease the helium abundance, y$^{+}$.  

\section{Reviewing the new sample} \label{Sample}

Applying the same criterion as AOS3, summarized in \S \ref{Recap}, table \ref{table:Cuts_AOS3} presents a summary of the cuts and flagged objects after reanalysis with the PFSD emissivities. The remaining 27 objects -- 16 qualifying and 11 flagged -- comprise our Final Dataset.  These are the objects for which the model is a good fit and which return physically meaningful parameter solutions, and they are used to determine Y$_p$ in the following section (\S \ref{Results}).  The new sample of this work is broadly similar to that found in AOS3.  The cuts have similar effects, with the main differences resulting from more objects satisfying the $\chi^{2}<4$ cut (35 here vs. 25 in AOS3).  Seven objects instead of two are excluded for unphysically large neutral hydrogen fractions ($\xi > 0.333$, $>$ 25\% neutral hydrogen), but most these added objects would have been excluded for the same reason in AOS3, if they had not already been excluded on the basis of $\chi^{2}$.  
One previously qualifying object is now flagged for $\xi > 0.01$, but otherwise, all previously qualifying and flagged objects retain their classification.  Additionally, three previously excluded objects are now qualifying, raising that sample from 14 to 16 objects, and two previously excluded objects are now flagged, raising that sample from 8 to 11 objects.

\begin{figure}
\centering  
\resizebox{\textwidth}{!}{\includegraphics{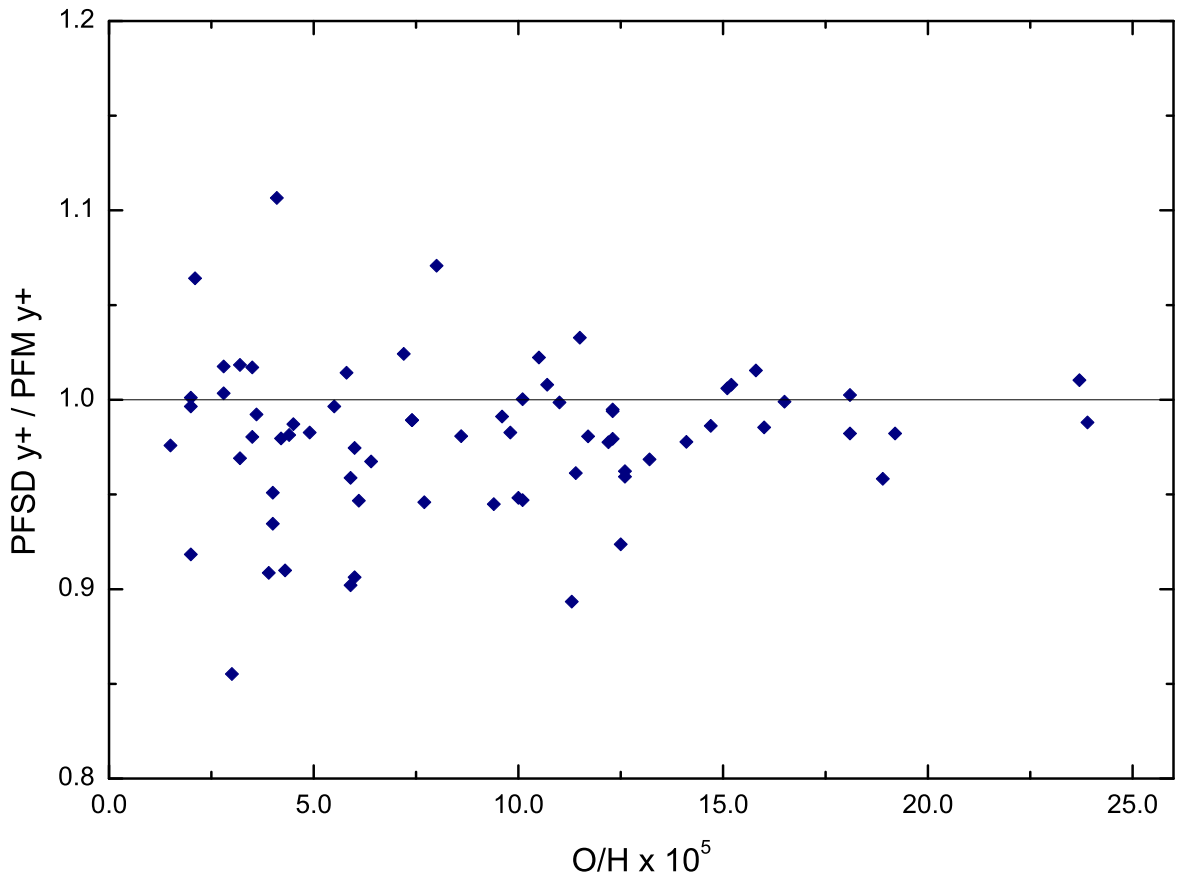}}
\caption{
Plot of the ratio of y$^{+}$ as determined in this work using the PFSD emissivities to that of AOS3 using the PFM emissivities.  This ratio is plotted versus O/H for the 70 objects for which He~I $\lambda$4026 was detected.  There is a clear trend to lower abundance values with the new PFSD emissivities.  Additionally, since metallicity is negatively correlated with temperature, the ratio tends further away from unity with decreasing abundance in accordance with the increasing divergence in the emissivity ratio as the temperature increases (see figure \ref{EmissivityRelative}).
}
\label{y+ratio}
\end{figure}

Each of the objects comprising the Final Dataset, with the results of their best-fit solutions and uncertainties, are presented in table \ref{table:GTO}, with the $\chi^2$ contributions for each of the six helium and three hydrogen lines given in table \ref{table:Chi}.  The best-fit solutions and uncertainties found in this work are broadly similar to those in AOS3.  Since the model is unchanged except for the emissivities, this is as expected.  The emissivities did, however, increase for all six lines, with the strong He~I lines $\lambda\lambda$5876 \& 6678 showing larger increases.  Correspondingly, compared to AOS3, y$^{+}$ generally decreases across the dataset, as can be seen in figure \ref{y+ratio}.  The average value of y$^{+}$ for the 13 mutual qualifying objects decreases by 1.6\%, while taken over the objects of the full qualifying dataset, y$^{+}$ decreases by 3.1\%.  This additional decrease in y$^{+}$ results from two of the three added objects (SBS~1415+437~(No.~1)~3 \& SBS~1030+583) having the lower abundances in the dataset.  On average the uncertainties on y$^{+}$ in this work increase by 1.5\% for the qualifying dataset compared to AOS3.  Looking at only the 13 mutual qualifying objects shows a 11\% increase in the average uncertainty on y$^{+}$.

Table \ref{table:Chi} shows which emission lines are making the strongest contribution to the $\chi^{2}$.  It could be hoped that there may be information from the Final Dataset which allows us to better understand why so many spectra fail our $\chi^{2}$ cut. For the qualifying set, in 10 of the 16 spectra, H$\gamma$ is making the strongest contribution to the $\chi^{2}$. For the other 6 spectra, the largest contributor is a He line.  Although H$\gamma$ appears to be the most discrepant line the most often, half the time (5 objects) the line is weak relative to the solution and half the time (5 objects) the line is strong relative to the solution.  The H$\gamma$ line is most influenced by reddening, underlying absorption, and collisional excitation.  Presently, there is no clear cause as to why H$\gamma$ is making the strongest contribution to the $\chi^{2}$.

\begin{figure}
\centering  
\resizebox{\textwidth}{!}{\includegraphics{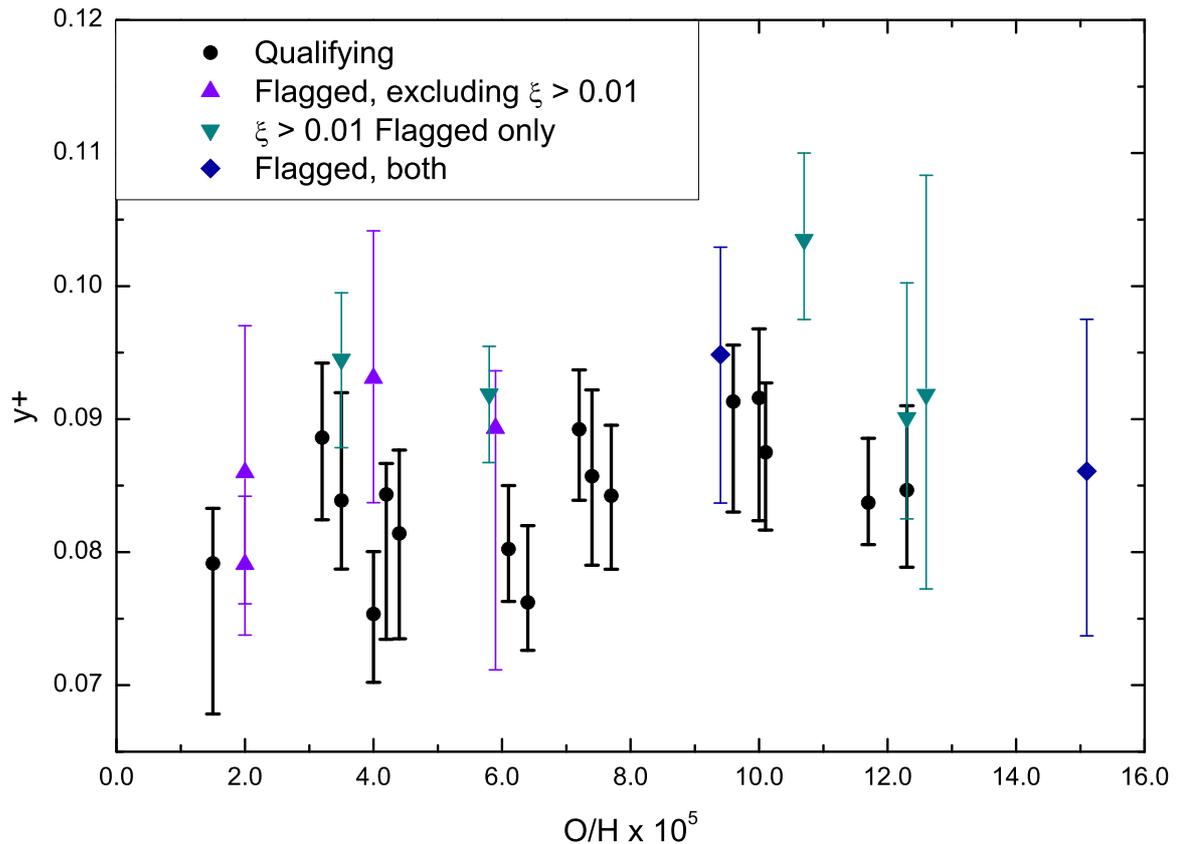}}
\caption{
Plot of y$^{+}$ vs O/H for the 27 objects meeting the prescribed reliability standards.  The upward triangles signify points flagged for large outlier values in optical depth or underlying absorption.  The downward triangles signify points flagged for large neutral hydrogen fractions.  The diamonds signify points flagged for both large outlier values in optical depth or underlying absorption and for large neutral hydrogen fractions. 
}
\label{y+OH}
\end{figure}

Figure \ref{y+OH} presents the derived y$^{+}$ values as a function of O/H.  The eleven objects flagged for large outlier values of $\tau$, a$_H$, a$_He$, and $\xi$ are highlighted with different symbols in figure \ref{y+OH}.  These flagged data points show a possible systematic shift to larger values of y$^{+}$; the average value of y$^{+}$ for the flagged objects is 8\% higher than that of the unflagged objects in the Final Dataset.

\begin{landscape}
\begin{table}[b!]\footnotesize
\centering
\vskip .1in
\begin{tabular}{lccccccccc}
\hline
\hline
Object 			& He$^+$/H$^+$     		   &  n$_e$     		      & a$_{He}$     		   & $\tau$     			& T$_e$     		   & C(H$\beta$)     	  	& a$_H$		     & $\xi$ $\times$ 10$^4$ 			& $\chi^2$ \\
\hline
\multicolumn{10}{c}{Final Dataset Not Flagged (Qualifying)} \\
\hline
I~Zw~18~SE~1    & 0.07916 $^{+0.00413}_{-0.01133}$ &       1 $^{+    431}_{-      1}$ &  0.22 $^{+ 0.19}_{- 0.22}$ &  0.56 $^{+ 0.69}_{- 0.56}$ & 18,330 $^{+2350}_{-3040}$ &  0.01 $^{+ 0.02}_{- 0.01}$ &  3.79 $^{+ 0.69}_{- 0.71}$ &        0 $^{+      31}_{-       0}$ &  0.3 \\
SBS~0940+544~2  & 0.08861 $^{+0.00561}_{-0.00617}$ &      52 $^{+    212}_{-     52}$ &  0.57 $^{+ 0.18}_{- 0.21}$ &  0.27 $^{+ 0.62}_{- 0.27}$ & 17,870 $^{+2050}_{-2310}$ &  0.02 $^{+ 0.05}_{- 0.02}$ &  2.88 $^{+ 1.10}_{- 1.41}$ &       35 $^{+     122}_{-      35}$ &  0.3 \\
Tol~65          & 0.08389 $^{+0.00810}_{-0.00516}$ &     639 $^{+    480}_{-    446}$ &  0.76 $^{+ 0.12}_{- 0.11}$ &  0.00 $^{+ 0.71}_{- 0.00}$ & 16,060 $^{+2640}_{-2410}$ &  0.00 $^{+ 0.05}_{- 0.00}$ &  5.67 $^{+ 0.34}_{- 1.11}$ &      186 $^{+     663}_{-     186}$ &  0.8 \\
SBS~1415+437~(No.~1)~3 & 0.07536 $^{+0.00467}_{-0.00515}$ &     148 $^{+    484}_{-    148}$ &  0.23 $^{+ 0.11}_{- 0.12}$ &  1.01 $^{+ 0.57}_{- 0.89}$ & 14,560 $^{+1840}_{-2380}$ &  0.13 $^{+ 0.02}_{- 0.04}$ &  0.02 $^{+ 0.75}_{- 0.02}$ &        0 $^{+     314}_{-       0}$ &  2.0 \\
SBS~1415+437~(No.~2) & 0.08434 $^{+0.00233}_{-0.01090}$ &       1 $^{+    587}_{-      1}$ &  0.51 $^{+ 0.07}_{- 0.15}$ &  1.30 $^{+ 0.63}_{- 1.25}$ & 15,590 $^{+1620}_{-2260}$ &  0.00 $^{+ 0.02}_{- 0.00}$ &  3.43 $^{+ 0.66}_{- 0.73}$ &        0 $^{+      73}_{-       0}$ &  1.3 \\
HS~1442+4250    & 0.08140 $^{+0.00628}_{-0.00792}$ &       1 $^{+    411}_{-      1}$ &  0.34 $^{+ 0.38}_{- 0.34}$ &  0.45 $^{+ 0.85}_{- 0.45}$ & 14,920 $^{+2390}_{-2530}$ &  0.03 $^{+ 0.04}_{- 0.03}$ &  0.00 $^{+ 1.15}_{- 0.00}$ &      142 $^{+    1290}_{-     142}$ &  1.8 \\
Mrk~209         & 0.08024 $^{+0.00477}_{-0.00395}$ &     121 $^{+    197}_{-    121}$ &  0.27 $^{+ 0.10}_{- 0.10}$ &  0.16 $^{+ 0.74}_{- 0.16}$ & 16,000 $^{+2030}_{-1890}$ &  0.00 $^{+ 0.02}_{- 0.00}$ &  2.61 $^{+ 0.66}_{- 1.06}$ &        0 $^{+      46}_{-       0}$ &  0.2 \\
SBS~1030+583    & 0.07623 $^{+0.00576}_{-0.00361}$ &     188 $^{+    236}_{-    188}$ &  0.19 $^{+ 0.10}_{- 0.08}$ &  0.00 $^{+ 0.85}_{- 0.00}$ & 14,550 $^{+1830}_{-1970}$ &  0.00 $^{+ 0.02}_{- 0.00}$ &  1.40 $^{+ 0.25}_{- 0.40}$ &        0 $^{+     112}_{-       0}$ &  1.3 \\
Mrk~71~(No.~1)  & 0.08924 $^{+0.00446}_{-0.00535}$ &       1 $^{+    324}_{-      1}$ &  0.56 $^{+ 0.15}_{- 0.15}$ &  2.14 $^{+ 0.60}_{- 0.64}$ & 14,950 $^{+1490}_{-2060}$ &  0.08 $^{+ 0.04}_{- 0.05}$ &  1.47 $^{+ 2.03}_{- 1.47}$ &      134 $^{+     989}_{-     134}$ &  1.5 \\
SBS~0917+527    & 0.08571 $^{+0.00648}_{-0.00669}$ &       1 $^{+    323}_{-      1}$ &  0.12 $^{+ 0.12}_{- 0.11}$ &  0.00 $^{+ 0.71}_{- 0.00}$ & 13,180 $^{+1300}_{-2160}$ &  0.04 $^{+ 0.05}_{- 0.04}$ &  0.50 $^{+ 0.63}_{- 0.50}$ &      566 $^{+    6400}_{-     566}$ &  0.8 \\
SBS~1152+579    & 0.08424 $^{+0.00530}_{-0.00552}$ &     220 $^{+    564}_{-    220}$ &  0.33 $^{+ 0.10}_{- 0.09}$ &  2.40 $^{+ 0.94}_{- 1.00}$ & 14,110 $^{+2040}_{-2390}$ &  0.19 $^{+ 0.06}_{- 0.06}$ &  1.11 $^{+ 1.31}_{- 1.11}$ &      297 $^{+    3180}_{-     297}$ &  3.2 \\
SBS~1054+365    & 0.09132 $^{+0.00425}_{-0.00830}$ &       1 $^{+    493}_{-      1}$ &  0.47 $^{+ 0.19}_{- 0.25}$ &  0.54 $^{+ 0.63}_{- 0.54}$ & 12,530 $^{+1670}_{-1980}$ &  0.00 $^{+ 0.06}_{- 0.00}$ &  2.84 $^{+ 0.56}_{- 0.99}$ &     1360 $^{+    8090}_{-    1364}$ &  0.6 \\
SBS~0926+606    & 0.09160 $^{+0.00518}_{-0.00924}$ &       1 $^{+    558}_{-      1}$ &  0.46 $^{+ 0.13}_{- 0.17}$ &  0.23 $^{+ 0.66}_{- 0.23}$ & 12,820 $^{+1550}_{-2250}$ &  0.01 $^{+ 0.06}_{- 0.01}$ &  0.59 $^{+ 0.58}_{- 0.59}$ &     1210 $^{+   12,450}_{-    1210}$ &  0.9 \\
Mrk~59          & 0.08751 $^{+0.00521}_{-0.00585}$ &      95 $^{+    501}_{-     95}$ &  0.54 $^{+ 0.08}_{- 0.08}$ &  0.38 $^{+ 0.56}_{- 0.38}$ & 14,950 $^{+1730}_{-2380}$ &  0.10 $^{+ 0.04}_{- 0.06}$ &  1.88 $^{+ 1.00}_{- 0.76}$ &       77 $^{+    1190}_{-      77}$ &  1.0 \\
SBS~1135+581    & 0.08373 $^{+0.00483}_{-0.00315}$ &     374 $^{+    681}_{-    374}$ &  0.40 $^{+ 0.06}_{- 0.05}$ &  0.00 $^{+ 0.60}_{- 0.00}$ & 12,020 $^{+1790}_{-1870}$ &  0.10 $^{+ 0.03}_{- 0.04}$ &  3.55 $^{+ 0.58}_{- 0.55}$ &        0 $^{+    4100}_{-       0}$ &  1.8 \\
HS~0924+3821    & 0.08466 $^{+0.00635}_{-0.00579}$ &       1 $^{+    797}_{-      1}$ &  0.35 $^{+ 0.16}_{- 0.14}$ &  0.34 $^{+ 0.76}_{- 0.34}$ & 12,250 $^{+1320}_{-2380}$ &  0.16 $^{+ 0.02}_{- 0.07}$ &  2.22 $^{+ 0.92}_{- 0.49}$ &        0 $^{+   10,410}_{-       0}$ &  0.4 \\
\hline
\multicolumn{10}{c}{Final Dataset with Flags} \\
\hline
SBS~0335-052E1  & 0.08598 $^{+0.01104}_{-0.00987}$ &     104 $^{+    366}_{-    104}$ &  0.20 $^{+ 0.19}_{- 0.19}$ &  5.14 $^{+ 1.00}_{- 0.96}$ & 18,120 $^{+2560}_{-2800}$ &  0.05 $^{+ 0.08}_{- 0.05}$ &  2.29 $^{+ 1.52}_{- 1.67}$ &       39 $^{+     245}_{-      39}$ &  0.5 \\
SBS~0335-052E2  & 0.07909 $^{+0.00512}_{-0.00533}$ &     314 $^{+    238}_{-    198}$ &  0.46 $^{+ 0.09}_{- 0.09}$ &  4.12 $^{+ 0.68}_{- 0.84}$ & 18,960 $^{+2570}_{-3160}$ &  0.00 $^{+ 0.05}_{- 0.00}$ &  3.61 $^{+ 0.52}_{- 1.13}$ &       15 $^{+      68}_{-      15}$ &  3.1 \\
UGC~4483        & 0.09449 $^{+0.00501}_{-0.00663}$ &       1 $^{+    284}_{-      1}$ &  0.35 $^{+ 0.09}_{- 0.08}$ &  0.49 $^{+ 0.54}_{- 0.49}$ & 14,270 $^{+1610}_{-2040}$ &  0.09 $^{+ 0.03}_{- 0.05}$ &  0.00 $^{+ 0.57}_{- 0.00}$ &      651 $^{+    3850}_{-     486}$ &  1.7 \\
HS~0122+0743    & 0.09308 $^{+0.01107}_{-0.00936}$ &     190 $^{+    337}_{-    190}$ &  1.12 $^{+ 0.27}_{- 0.24}$ &  0.26 $^{+ 0.84}_{- 0.26}$ & 17,350 $^{+2300}_{-2500}$ &  0.02 $^{+ 0.08}_{- 0.02}$ &  5.43 $^{+ 1.11}_{- 2.18}$ &       79 $^{+     325}_{-      79}$ &  0.5 \\
SBS~1331+493    & 0.09187 $^{+0.00360}_{-0.00513}$ &       1 $^{+    195}_{-      1}$ &  0.18 $^{+ 0.13}_{- 0.12}$ &  1.38 $^{+ 0.31}_{- 0.65}$ & 13,770 $^{+1620}_{-1370}$ &  0.00 $^{+ 0.02}_{- 0.00}$ &  0.13 $^{+ 0.74}_{- 0.13}$ &      760 $^{+    1450}_{-     585}$ &  3.6 \\
UM~461          & 0.08932 $^{+0.00431}_{-0.01817}$ &      25 $^{+    650}_{-     25}$ &  0.71 $^{+ 0.30}_{- 0.47}$ &  2.20 $^{+ 0.71}_{- 1.49}$ & 18,200 $^{+2320}_{-3060}$ &  0.00 $^{+ 0.06}_{- 0.00}$ &  7.68 $^{+ 0.97}_{- 1.59}$ &       30 $^{+     157}_{-      30}$ &  1.5 \\
HS~0811+4913    & 0.09485 $^{+0.00808}_{-0.01116}$ &     170 $^{+    641}_{-    170}$ &  1.18 $^{+ 0.40}_{- 0.46}$ &  0.69 $^{+ 0.82}_{- 0.69}$ & 14,780 $^{+1760}_{-2230}$ &  0.00 $^{+ 0.04}_{- 0.00}$ &  9.46 $^{+ 1.47}_{- 2.19}$ &      434 $^{+    1780}_{-     347}$ &  0.9 \\
HS~1214+3801    & 0.10349 $^{+0.00650}_{-0.00600}$ &     414 $^{+    406}_{-    238}$ &  0.82 $^{+ 0.16}_{- 0.16}$ &  0.00 $^{+ 0.25}_{- 0.00}$ & 13,930 $^{+1630}_{-1580}$ &  0.16 $^{+ 0.04}_{- 0.04}$ &  5.71 $^{+ 1.34}_{- 1.33}$ &     1410 $^{+    4590}_{-    1011}$ &  2.6 \\
UM~439          & 0.09009 $^{+0.01016}_{-0.00759}$ &     543 $^{+    635}_{-    393}$ &  0.05 $^{+ 0.19}_{- 0.05}$ &  2.88 $^{+ 1.17}_{- 1.41}$ & 13,770 $^{+2170}_{-2560}$ &  0.12 $^{+ 0.06}_{- 0.06}$ &  2.18 $^{+ 1.36}_{- 1.25}$ &      968 $^{+    9520}_{-     822}$ &  1.0 \\
SBS~1533+574B   & 0.09185 $^{+0.01648}_{-0.01461}$ &     847 $^{+   2906}_{-    697}$ &  0.42 $^{+ 0.18}_{- 0.18}$ &  0.85 $^{+ 1.69}_{- 0.85}$ & 12,440 $^{+2280}_{-2050}$ &  0.23 $^{+ 0.10}_{- 0.09}$ &  3.07 $^{+ 0.99}_{- 1.07}$ &     4110 $^{+   36,270}_{-    3720}$ &  1.3 \\
UM~238          & 0.08609 $^{+0.01142}_{-0.01238}$ &    1810 $^{+   6620}_{-   1080}$ &  1.13 $^{+ 0.69}_{- 0.73}$ &  0.37 $^{+ 1.26}_{- 0.37}$ & 12,250 $^{+2350}_{-2280}$ &  0.15 $^{+ 0.07}_{- 0.07}$ & 17.05 $^{+ 2.92}_{- 3.19}$ &     2530 $^{+   33,990}_{-    2300}$ &  1.0 \\
\hline
\end{tabular}
\caption{Physical parameters and He$^+$/H$^+$ abundance solutions of Final Dataset}
\label{table:GTO}
\end{table}
\end{landscape}

\begin{landscape}
\begin{table}[b!]\footnotesize
\centering
\vskip .1in
\begin{tabular}{lccccccccc}
\hline
\hline
Object 			& He~$\lambda$3889     		   &  He~$\lambda$4026     		      & He~$\lambda$4471     		   & He~$\lambda$5876     			& He~$\lambda$6678     		   & He~$\lambda$7065     	  	& H$\alpha$		     & H$\gamma$ 			& H$\delta$ \\
\hline
\multicolumn{10}{c}{Final Dataset Not Flagged (Qualifying)} \\
\hline
I~Zw~18~SE~1		&	0.078	&	-0.127	&	0.096	&	-0.104	&	0.047	&	0.055	&	0.112	&	0.408	&	-0.254	\\
SBS~0940+544~2		&	0.065	&	0.002	&	-0.022	&	0.104	&	-0.247	&	0.024	&	-0.055	&	-0.401	&	0.046	\\
Tol~65			&	0.115	&	-0.124	&	0.132	&	-0.451	&	0.563	&	0.047	&	-0.069	&	-0.281	&	-0.046	\\
SBS~1415+437~(No.~1)~3	&	0.288	&	-0.197	&	0.204	&	0.258	&	-0.943	&	0.104	&	0.146	&	-0.439	&	-0.308	\\
SBS~1415+437~(No.~2)	&	0.276	&	-0.136	&	0.479	&	-0.328	&	0.121	&	0.130	&	0.666	&	0.442	&	-0.435	\\
HS~1442+4250		&	0.351	&	-0.351	&	0.224	&	-0.257	&	-0.154	&	0.188	&	-0.192	&	-0.337	&	-0.748	\\
Mrk~209			&	0.017	&	-0.006	&	0.013	&	-0.019	&	0.000	&	0.008	&	0.112	&	0.410	&	-0.168	\\
SBS~1030+583		&	0.097	&	-0.463	&	0.783	&	-0.416	&	0.183	&	0.136	&	0.309	&	0.048	&	-0.096	\\
Mrk~71~(No.~1)		&	0.058	&	0.006	&	-0.130	&	-0.356	&	0.675	&	0.049	&	0.006	&	0.776	&	-0.467	\\
SBS~0917+527		&	0.183	&	-0.129	&	0.135	&	-0.110	&	-0.184	&	0.239	&	-0.075	&	-0.452	&	-0.064	\\
SBS~1152+579		&	0.185	&	-0.028	&	0.097	&	0.325	&	-0.806	&	0.048	&	-0.140	&	-1.458	&	0.255	\\
SBS~1054+365		&	0.192	&	-0.085	&	-0.196	&	-0.187	&	0.289	&	0.145	&	-0.089	&	-0.453	&	-0.111	\\
SBS~0926+606		&	0.061	&	-0.219	&	0.536	&	-0.169	&	-0.052	&	0.043	&	-0.038	&	-0.621	&	0.230	\\
Mrk~59			&	-0.105	&	0.127	&	-0.296	&	0.436	&	-0.315	&	-0.083	&	0.103	&	0.532	&	0.028	\\
SBS~1135+581		&	-0.014	&	-0.152	&	0.433	&	-0.607	&	0.633	&	0.094	&	0.262	&	0.807	&	-0.221	\\
HS~0924+3821		&	0.032	&	-0.148	&	0.153	&	-0.186	&	0.466	&	0.053	&	0.050	&	0.313	&	-0.159	\\
\hline
\multicolumn{10}{c}{Final Dataset with Flags} \\
\hline
SBS~0335-052E1	&	0.181	&	-0.088	&	0.058	&	-0.128	&	-0.058	&	0.130	&	-0.096	&	-0.357	&	-0.143	\\
SBS~0335-052E2	&	0.103	&	-0.212	&	0.492	&	-1.048	&	1.164	&	0.116	&	-0.109	&	-0.457	&	-0.084	\\
UGC~4483	&	0.247	&	-0.029	&	-0.015	&	0.233	&	-0.597	&	0.069	&	-0.187	&	-0.726	&	-0.243	\\
HS~0122+0743	&	0.049	&	0.045	&	-0.130	&	0.262	&	-0.450	&	0.007	&	-0.051	&	-0.470	&	0.106	\\
SBS~1331+493	&	0.266	&	-0.190	&	0.024	&	-1.029	&	0.946	&	0.145	&	-0.157	&	-0.841	&	-0.145	\\
UM~461		&	-0.267	&	0.423	&	-0.395	&	0.532	&	-0.318	&	-0.153	&	0.084	&	-0.321	&	0.509	\\
HS~0811+4913	&	-0.030	&	0.037	&	-0.065	&	0.040	&	-0.012	&	-0.012	&	-0.013	&	-0.825	&	0.400	\\
HS~1214+3801	&	-0.459	&	-0.201	&	0.770	&	0.003	&	-0.583	&	0.208	&	0.183	&	-0.773	&	0.839	\\
UM~439		&	0.072	&	-0.219	&	0.439	&	-0.098	&	-0.093	&	0.049	&	-0.057	&	-0.794	&	0.227	\\
SBS~1533+574B	&	0.107	&	-0.191	&	0.354	&	0.089	&	-0.645	&	0.034	&	-0.055	&	-0.812	&	0.244	\\
UM~238		&	-0.072	&	0.067	&	-0.128	&	-0.154	&	0.550	&	-0.007	&	0.038	&	0.739	&	-0.247	\\
\hline
\end{tabular}
\caption{
$\sqrt{\chi^{2}}$ contributions by line for the Final Dataset.  
Positive values indicate the predicted flux is greater than the measured flux, see eq. \ref{eq:X2}. 
}
\label{table:Chi}
\end{table}
\end{landscape}

\section{Results from the Final Dataset} \label{Results}

The primary goal of this work, the primordial helium abundance (mass fraction), Y$_{p}$, can now be calculated for several subsets of the Final Dataset.  A regression of Y, the helium mass fraction, versus O/H, the oxygen to hydrogen mass fraction, is used to extrapolate to the primordial value\footnote{This work takes $Z=20(O/H)$ such that $Y=\frac{4y(1-20(O/H))}{1+4y}$}.  The O/H values are taken directly from ITS07.  

\begin{figure}
\resizebox{\textwidth}{!}{\includegraphics{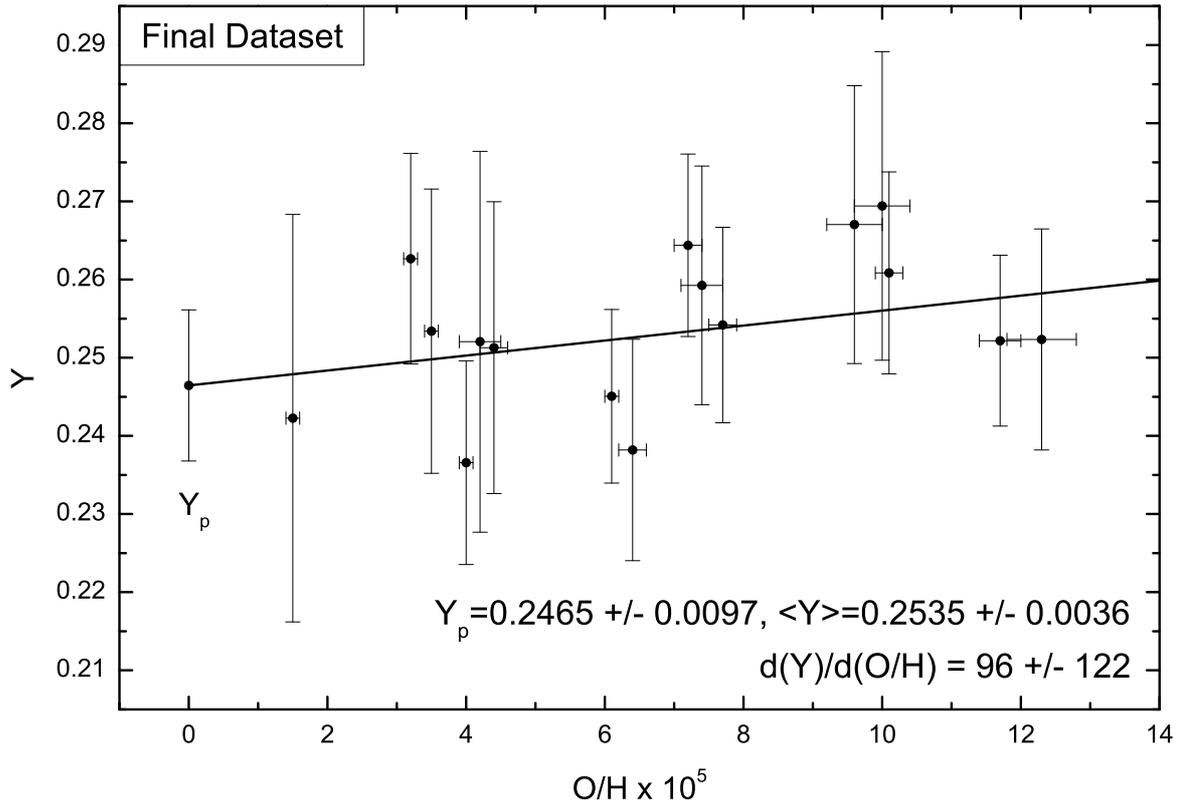}}
\caption{
Helium abundance (mass fraction) versus oxygen to hydrogen ratio regression calculating the primordial helium abundance.
}
\label{Y_OH}
\end{figure}

Because it minimizes confounding systematic effects, our preferred dataset is the 16 qualifying points.  
The relevant values for its regression are given in table \ref{table:PH}.  The regression yields,
\beq
Y_p = 0.2465 \pm 0.0097,
\label{eq:Yp}
\eeq
with a slope of 96 $\pm$ 122 and a $\chi^2$ of 6.1.  The result is shown in figure \ref{Y_OH}.  
This result for $Y_p$ agrees well with the Planck value of $Y_p = 0.2485 \pm 0.0002$. AOS3 determined $Y_p = 0.2534 \pm 0.0083$ with a slope of 54 $\pm$ 102.  Given their large uncertainties, AOS3's result is in agreement with the newer result.  Despite the slightly increased dataset (from 14 to 16), this work finds a slightly larger uncertainty.  This is directly a result of the larger variance in the dataset, as is clearly evidenced by the increase in the $\chi^2$ of the fit from 2.9 to 6.1.  However, the dispersion in the dataset is still well captured by the individual uncertainties.  The inclusion of two additional, lower abundance objects to the qualifying dataset, combined with the increase in the PFSD emissivities relative to PFM, particularly for the strong lines He~I $\lambda\lambda$5876 \& 6678, results in the decrease in the abundances and $Y_p$.

As the O/H domain is limited, an estimate of Y$_p$ using the mean value is justified and gives,
\beq
Y_p = 0.2535 \pm 0.0036.
\eeq
This is not significantly different from the result of the regression fit; however, the uncertainty is decreased by more than a factor of two.  

Including the flagged objects raises the intercept and reduces the uncertainty to $0.2509 \pm 0.0071$ with a slope of 105 $\pm$ 90.  The reduced uncertainty is a result of the increased number of points in the regression, and the possible systematic bias toward larger y$^{+}$ within the flagged dataset raises the intercept.  \citet{os04} restricted the metallicity baseline to $O/H = 9.2 \times 10^{-5}$.  Adopting the same metallicity cut with the dataset of this work decreases the intercept slightly to $0.2441 \pm 0.0147$.  Using all 93 observations included in their HeBCD sample, ITS07 determined $Y_p = 0.2516 \pm 0.0011$.  Their much smaller uncertainty is achieved primarily though the use of the full sample of observations.  In a more recent analysis using their HeBCD sample and observations from the SDSS and VLT, \citet{isg13} find $Y_p = 0.254 \pm 0.003$.

Finally, we also include analysis of the recently discovered extremely metal deficient dwarf galaxy Leo~P \citep{leop}.  Because of Leo~P's low metallicity it is particularly valuable in determining Y$_p$, and its best fit solution and regression parameters are given in table \ref{table:LeoP}.  Leo~P satisfies all of the same quality and reliability criteria as our qualifying dataset, and including it in a regression with the qualifying points returns an intercept of $0.2463 \pm 0.0090$ and a slope of 97 $\pm$ 115.  Leo~P agrees very well with the regression determined by the qualifying dataset alone (eq. \ref{eq:Yp}), and as a result, the regression is essentially unchanged by the addition of Leo~P, except for a small decrease in the intercept's uncertainty.
Table \ref{table:Yp's} summarizes the calculated regression Y$_p$ and slope, as well as the mean, $<Y>$, for several subsets of the Final Dataset found in this work.

\begin{table}[ht!]
\centering
\vskip .1in
\begin{tabular}{lcccc}
\hline\hline
Object & 	He$^+$/H$^+$ 	      & He$^{++}$/H$^+$     & Y 		  & O/H $\times$ 10$^5$ \\
\hline
\multicolumn{5}{c}{Final Dataset Not Flagged (Qualifying)} \\
\hline
I~Zw~18~SE~1		&	0.07916	$\pm$	0.01133	&	0.0008	$\pm$	0.0008	&	0.2423	$\pm$	0.0261	&		1.5	$\pm$	0.1	\\
SBS~0940+544~2		&	0.08861	$\pm$	0.00617	&	0.0005	$\pm$	0.0005	&	0.2627	$\pm$	0.0135	&		3.2	$\pm$	0.1	\\
Tol~65			&	0.08389	$\pm$	0.00810	&	0.0010	$\pm$	0.0010	&	0.2534	$\pm$	0.0182	&		3.5	$\pm$	0.1	\\
SBS~1415+437~(No.~1)~3	&	0.07536	$\pm$	0.00515	&	0.0022	$\pm$	0.0022	&	0.2366	$\pm$	0.0130	&		4.0	$\pm$	0.1	\\
SBS~1415+437~(No.~2)	&	0.08434	$\pm$	0.01090	&	0.0000			&	0.2521	$\pm$	0.0244	&		4.2	$\pm$	0.3	\\
HS~1442+4250		&	0.08140	$\pm$	0.00792	&	0.0026	$\pm$	0.0026	&	0.2513	$\pm$	0.0187	&		4.4	$\pm$	0.2	\\
Mrk~209			&	0.08024	$\pm$	0.00477	&	0.0010	$\pm$	0.0010	&	0.2451	$\pm$	0.0111	&		6.1	$\pm$	0.1	\\
SBS~1030+583		&	0.07623	$\pm$	0.00576	&	0.0021	$\pm$	0.0021	&	0.2382	$\pm$	0.0142	&		6.4	$\pm$	0.2	\\
Mrk~71~(No.~1)		&	0.08924	$\pm$	0.00535	&	0.0008	$\pm$	0.0008	&	0.2644	$\pm$	0.0117	&		7.2	$\pm$	0.2	\\
SBS~0917+527		&	0.08571	$\pm$	0.00669	&	0.0020	$\pm$	0.0020	&	0.2593	$\pm$	0.0153	&		7.4	$\pm$	0.3	\\
SBS~1152+579		&	0.08424	$\pm$	0.00552	&	0.0011	$\pm$	0.0011	&	0.2542	$\pm$	0.0125	&		7.7	$\pm$	0.2	\\
SBS~1054+365		&	0.09132	$\pm$	0.00830	&	0.0000			&	0.2670	$\pm$	0.0178	&		9.6	$\pm$	0.4	\\
SBS~0926+606		&	0.09160	$\pm$	0.00924	&	0.0008	$\pm$	0.0008	&	0.2694	$\pm$	0.0197	&		10.0	$\pm$	0.4	\\
Mrk~59			&	0.08751	$\pm$	0.00585	&	0.0010	$\pm$	0.0010	&	0.2609	$\pm$	0.0129	&		10.1	$\pm$	0.2	\\
SBS~1135+581		&	0.08373	$\pm$	0.00483	&	0.0008	$\pm$	0.0008	&	0.2522	$\pm$	0.0109	&		11.7	$\pm$	0.3	\\
HS~0924+3821		&	0.08466	$\pm$	0.00635	&	0.0000			&	0.2523	$\pm$	0.0142	&		12.3	$\pm$	0.5	\\
\hline
\multicolumn{5}{c}{Final Dataset with Flags} \\
\hline
SBS~0335-052E1	&	0.08598	$\pm$	0.01104	&	0.0023	$\pm$	0.0023	&	0.2609	$\pm$	0.0246	&	2.0	$\pm$	0.1	\\
SBS~0335-052E2	&	0.07909	$\pm$	0.00533	&	0.0024	$\pm$	0.0024	&	0.2458	$\pm$	0.0133	&	2.0	$\pm$	0.1	\\
UGC~4483	&	0.09449	$\pm$	0.00663	&	0.0000			&	0.2741	$\pm$	0.0140	&	3.5	$\pm$	0.1	\\
HS~0122+0743	&	0.09308	$\pm$	0.01107	&	0.0007	$\pm$	0.0007	&	0.2726	$\pm$	0.0235	&	4.0	$\pm$	0.1	\\
SBS~1331+493	&	0.09187	$\pm$	0.00513	&	0.0000			&	0.2684	$\pm$	0.0110	&	5.8	$\pm$	0.2	\\
UM~461		&	0.08932	$\pm$	0.01817	&	0.0000			&	0.2629	$\pm$	0.0394	&	5.9	$\pm$	0.3	\\
HS~0811+4913	&	0.09485	$\pm$	0.01116	&	0.0003	$\pm$	0.0003	&	0.2752	$\pm$	0.0234	&	9.4	$\pm$	0.2	\\
HS~1214+3801	&	0.10349	$\pm$	0.00650	&	0.0000			&	0.2921	$\pm$	0.0130	&	10.7	$\pm$	0.2	\\
UM~439		&	0.09009	$\pm$	0.01016	&	0.0000			&	0.2642	$\pm$	0.0219	&	12.3	$\pm$	0.3	\\
SBS~1533+574B	&	0.09185	$\pm$	0.01648	&	0.0009	$\pm$	0.0009	&	0.2698	$\pm$	0.0350	&	12.6	$\pm$	0.7	\\
UM~238		&	0.08609	$\pm$	0.01238	&	0.0000			&	0.2554	$\pm$	0.0273	&	15.1	$\pm$	0.5	\\
\hline
\end{tabular}
\caption{Primordial helium regression values}
\label{table:PH}
\end{table}

\begin{table}[ht!]
\centering
\vskip .1in
\begin{tabular}{lc}
\hline\hline
He$^+$/H$^+$			& 0.08202 $^{+0.00831}_{-0.00915}$ \\
n$_e$				& 1 $^{+   324}_{-   1}$ \\
a$_{He}$			& 0.45 $^{+ 0.45}_{- 0.41}$ \\
$\tau$				& 0.00 $^{+ 0.76}_{- 0.00}$ \\
T$_e$				& 17,440 $^{+1430}_{-3320}$ \\
C(H$\beta$)			& 0.10 $^{+ 0.03}_{- 0.08}$ \\
a$_H$				& 0.94 $^{+ 1.50}_{- 0.94}$ \\
$\xi$ $\times$ 10$^4$   	& 0 $^{+    151}_{-     0}$ \\
$\chi^2$			& 3.2 \\
O/H $\times$ 10$^5$		& 1.5 $\pm$ 0.1 \\
Y				& 0.2470 $\pm$ 0.0208 \\
\hline
\end{tabular}
\caption{Physical conditions, He$^+$/H$^+$ abundance solution, and regression values of Leo~P}
\label{table:LeoP}
\end{table}

\begin{table}[ht!]
\centering
\vskip .1in
\begin{tabular}{lcccc}
\hline\hline
Dataset							& N    	 & Y$_p$	 	& $dY/d(O/H)$	& $<Y>$ \\
\hline
Qualifying						& 16 	& 0.2465 $\pm$ 0.0097 	&  96 $\pm$ 122	& 0.2535 $\pm$ 0.0036 \\
Qualifying + Flagged, excluding $\xi>0.01$		& 20	& 0.2482 $\pm$ 0.0080	&  79 $\pm$ 106	& 0.2536 $\pm$ 0.0034 \\
Qualifying + $\xi>0.01$ Flagged only			& 21 	& 0.2505 $\pm$ 0.0084	& 108 $\pm$ 105	& 0.2585 $\pm$ 0.0032 \\
Qualifying + All Flagged				& 27	& 0.2509 $\pm$ 0.0071	& 105 $\pm$  90	& 0.2584 $\pm$ 0.0030 \\
Qualifying, $O/H < 9.2 \times 10^{-5}$ (AOS/AOS2)	& 11	& 0.2441 $\pm$ 0.0147	& 129 $\pm$ 251	& 0.2514 $\pm$ 0.0044 \\
Qualifying + Leo~P					& 17	& 0.2463 $\pm$ 0.0090	&  97 $\pm$ 115	& 0.2533 $\pm$ 0.0036 \\

\hline
\end{tabular}
\caption{Comparison of Y$_p$ for selected datasets}
\label{table:Yp's}
\end{table}

\section{Discussion} \label{Conclusion}

Given the central role emissivities play in H~II region analysis, we have updated our analysis to incorporate the new PFSD emissivities, which utilize the most recent atomic data.  Following the careful screening of AOS3 for goodness-of-fit, minimization of systematic bias, and physically meaningful results, we applied the MCMC method and determined the primordial helium abundance from a dataset of 16 objects.  Highlighting the robust screening, 13 of these objects are retained from the 14 objects of AOS3, and multiple regressions including flagged objects show strong agreement.  A general increase in the emissivities coupled with the addition of two lower abundance qualifying objects leads to a lower primordial helium abundance in good agreement with previous results and the Planck determination.
 
Higher quality, higher resolution spectra hold the most promise for significant future improvement in the primordial helium abundance determination.  Most directly, increased measurement precision more tightly constrains the solution, but the potential for more reliable objects and an increased dataset is an equally strong motivation.  Furthermore, as discussed in AOS, higher resolution spectra afford the chance to measure underlying Balmer absorption directly, while higher signal to noise offer the possibility of adding weaker Helium and Hydrogen lines to our analysis.  In general, an increase in the number of lines relative to the number of parameters reduces degeneracies and better defines and constrains the solution. 

In summary, the effect of the new emissivities is to bring our determination of the primordial helium abundance into better agreement with the value inferred by the CMB determined baryon density.  The consistency of our dataset and results supports the effectiveness of Monte Carlo methods and careful screening for quality and reliability.  The persistence of relatively large uncertainties is not unexpected given the uncertainties in the measurement and model involved.  While we believe that our results underscore the need for higher quality spectra, further work in establishing the root cause 
for the discrepancy between theory and data in the majority of observed objects is also warranted.

\acknowledgments

The work of KAO is supported in part by DOE grant DE-FG02-94ER-40823. 
EDS is grateful for partial support from the University of Minnesota.


\newpage
\begin{thebibliography}{}

\bibitem[Ade et al.(2013)]{planck} 
  P.~A.~R.~Ade {\it et al.}  [Planck Collaboration],
  arXiv:1303.5082 [astro-ph.CO].
  
\bibitem[Cyburt et al.(2002)]{cfo2}
  R.~H.~Cyburt, B.~D.~Fields and K.~A.~Olive,
  Astropart.\ Phys.\ \ {\bf 17}, 87  (2002)
  [astro-ph/0105397].

\bibitem[Walker et al.(1991)]{wssok} 
  T.~P.~Walker, G.~Steigman, D.~N.~Schramm, K.~A.~Olive and H.~S.~Kang,
  Astrophys.\ J.\  {\bf 376}, 51 (1991).

\bibitem[Olive, Steigman, \& Walker(2000)]{osw}
  K.~A.~Olive, G.~Steigman and T.~P.~Walker,
  Phys.\ Rept.\  {\bf 333}, 389 (2000)
  [arXiv:astro-ph/9905320].

\bibitem[Fields \& Sarkar(2008)]{fs} 
  B.~D.~Fields and S.~Sarkar,
  in K. Nakamura et al.
  J.\ Phys.\ G {\bf 37}, 075021 (2010)
 
\bibitem[Cyburt et al.(2001)]{cfo} 
  R.~H.~Cyburt, B.~D.~Fields and K.~A.~Olive,
  New Astron.\  {\bf 6}, 215 (2001)
  [arXiv:astro-ph/0102179].

\bibitem[Coc et al.(2002)]{coc} 
  A.~Coc, E.~Vangioni-Flam, M.~Casse and M.~Rabiet,
  Phys.\ Rev.\  D {\bf 65}, 043510 (2002)
  [arXiv:astro-ph/0111077].
  
  \bibitem[Cyburt et al.(2003)]{cfo3}
  R.~H.~Cyburt, B.~D.~Fields and K.~A.~Olive,
  Phys.\ Lett.\ B\ {\bf 567}, 227  (2003)
  [astro-ph/0302431].

\bibitem[Coc et al.(2004)]{coc2} 
  A.~Coc, E.~Vangioni-Flam, P.~Descouvemont, A.~Adahchour and C.~Angulo,
  Astrophys.\ J.\  {\bf 600}, 544 (2004)
  [arXiv:astro-ph/0309480].

\bibitem[Cyburt(2004)]{cyburt} 
  R.~H.~Cyburt,
  Phys.\ Rev.\  D {\bf 70}, 023505 (2004)
  [arXiv:astro-ph/0401091].

\bibitem[Descouvemont et al.(2004)]{coc3} 
  P.~Descouvemont, A.~Adahchour, C.~Angulo, A.~Coc and E.~Vangioni-Flam,
  Atomic Data and Nuclear Data Tables, {\bf 88}, 203 (2004)
  [arXiv:astro-ph/0407101].

\bibitem[Cuoco et al.(2004)]{cuoco} 
  A.~Cuoco, F.~Iocco, G.~Mangano, G.~Miele, O.~Pisanti and P.~D.~Serpico,
  Int.\ J.\ Mod.\ Phys.\  A {\bf 19}, 4431 (2004)
  [arXiv:astro-ph/0307213].

\bibitem[Serpico et al.(2004)]{serp} 
  P.~D.~Serpico, S.~Esposito, F.~Iocco, G.~Mangano, G.~Miele and O.~Pisanti,
  JCAP {\bf 0412}, 010 (2004)
  [arXiv:astro-ph/0408076].

\bibitem[Cyburt et al.(2008)]{cfo5} 
  R.~H.~Cyburt, B.~D.~Fields and K.~A.~Olive,
  JCAP {\bf 0811}, 012 (2008)
  [arXiv:0808.2818 [astro-ph]].

\bibitem[Coc et al.(2011)]{coc4} 
  A.~Coc, S.~Goriely, Y.~Xu, M.~Saimpert and E.~Vangioni,
  arXiv:1107.1117 [astro-ph.CO].
  
  \bibitem[Coc et al.(2013)]{cuv}
  A.~Coc, J.~-P.~Uzan and E.~Vangioni,
  arXiv:1307.6955 [astro-ph.CO].

\bibitem[Beringer et al.(2012)]{rpp} 
  J.~Beringer {\it et al.}  [Particle Data Group],
  Phys.\ Rev.\ D {\bf 86}, 010001 (2012).
  
\bibitem{deut}
  S.~Burles and D.~Tytler,
  Astrophys.\ J.\  {\bf 499}, 699 (1998)
  [arXiv:astro-ph/9712108];
  S.~Burles and D.~Tytler,
  Astrophys.\ J.\  {\bf 507}, 732 (1998)
  [arXiv:astro-ph/9712109];

\bibitem{cfos}
  R.~H.~Cyburt, B.~D.~Fields, K.~A.~Olive and E.~Skillman,
  Astropart.\ Phys.\  {\bf 23}, 313 (2005)
  [arXiv:astro-ph/0408033].
  
  \bibitem{cooke}
  R.~Cooke, M.~Pettini, R.~A.~Jorgenson, M.~T.~Murphy and C.~C.~Steidel,
  arXiv:1308.3240 [astro-ph.CO].

\bibitem[Peimbert \& Torres-Peimbert(1974)]{ptp74} 
  M.~Peimbert and S.~Torres-Peimbert,
  Astrophys.\ J.\  {\bf 193}, 327 (1974)

\bibitem[Olive \& Skillman(2001)]{os01}  
  K.~A.~Olive and E.~D.~Skillman,
  New Astron.\ {\bf 6}, 119 (2001)
  arXiv:astro-ph/0007081.

\bibitem[Olive \& Skillman(2004)]{os04}  
  K.~A.~Olive and E.~D.~Skillman,
  Astrophys.\ J.\  {\bf 617}, 29 (2004)
  [arXiv:astro-ph/0405588].

\bibitem[Izotov, Thuan, \& Stasi\'nska(2007)]{its07} 
  Y.~I.~Izotov, T.~X.~Thuan and G.~Stasi\'nska,
  Astrophys.\ J.\  {\bf 662}, 15 (2007)
  [arXiv:astro-ph/0702072] (ITS07).

\bibitem[Peimbert, Luridiana, \& Peimbert(2007)]{plp07}
  M.~Peimbert, V.~Luridiana and A.~Peimbert,
  Astrophys.\ J.\  {\bf 666}, 636 (2007)
  [arXiv:astro-ph/0701580].

\bibitem[Izotov, Thuan, \& Lipovetsky(1994)]{itl94}
  Y.~I.~Izotov, T.~X.~Thuan and V.~A.~Lipovetsky,
  Astrophys.\ J.\  {\bf 435}, 647 (1994).
  
\bibitem[Izotov, Thuan, \& Lipovetsky(1997)]{itl97}
   Y.~I.~Izotov, T.~X.~Thuan and V.~A.~Lipovetsky,
  Astrophys.\ J.\ Suppl.\  {\bf 108}, 1 (1997).

\bibitem[Peimbert, Peimbert, \& Ruiz(2000)]{ppr00}
  M.~Peimbert, A.~Peimbert and M.~T.~Ruiz,
  Astrophys.\ J.\  {\bf 541}, 688 (2000)
  [arXiv:astro-ph/0003154].

\bibitem[Aver et al.(2010)]{AOS} 
  E.~Aver, K.~A.~Olive and E.~D.~Skillman,
  JCAP {\bf 1005}, 003 (2010)
  [arXiv:1001.5218 [astro-ph.CO]] (AOS).

\bibitem[Aver et al.(2011)]{AOS2}
  E.~Aver, K.~A.~Olive and E.~D.~Skillman,
  JCAP {\bf 1103}, 043 (2011)
  [arXiv:1012.2385 [astro-ph.CO]] (AOS2).

\bibitem[Aver et al.(2012)]{AOS3}
  E.~Aver, K.~A.~Olive and E.~D.~Skillman,
  JCAP {\bf 1204}, 004 (2012)
  [arXiv:1112.3713 [astro-ph.CO]] (AOS3).
  
\bibitem[Smits(1991)]{smits91}
  D.~P.~Smits, MNRAS {\bf 251}, 316 (1991).
  
\bibitem[Smits(1996)]{smits96}
  D.~P.~Smits, MNRAS {\bf 278}, 683 (1996).
  
\bibitem[Brocklehurst(1972)]{brock}
  M.~Brocklehurst, MNRAS {\bf 157}, 211 (1972).
  
\bibitem[Sawey \& Berrington(1993)]{sb93}
  P.~M.~J.~Sawey \& K.~A.~Berrington, At. Data Nucl. Data Tables {\bf 55},
  82 (1993).
  
\bibitem[Benjamin, Skillman, \& Smits(1999)]{bss99}
  R.~A.~Benjamin, E.~D.~Skillman and D.~P.~Smits,
  Astrophys.\ J.\  {\bf 514}, 307 (1999)
  [arXiv:astro-ph/9810087] (BSS).

\bibitem[Porter et al.(2005)]{por05} 
  R.~L.~Porter, R.~P.~Bauman, G.~J.~Ferland and K.~B.~MacAdam,
  Astrophys.\ J.\  {\bf 622}, L73 (2005)
  [arXiv:astro-ph/0502224].

\bibitem[Porter, Ferland, \& MacAdam(2007)]{pfm07} 
  R.~L.~Porter, G.~J.~Ferland and K.~B.~MacAdam,
  Astrophys.\ J.\  {\bf 657}, 327 (2007)
  [arXiv:astro-ph/0611579] (PFM).

\bibitem[Porter, Ferland, Storey, \& Detisch(2012)]{pfsd} 
  R.~L.~Porter, G.~J.~Ferland, P.~J.~Storey and M.~J.~Detisch,
  Mon.\ Not.\ Roy.\ Astron.\ Soc.\  {\bf 425}, L28 (2012)
  arXiv:1206.4115 [astro-ph.CO] (PFSD).

\bibitem[Porter, Ferland, Storey, \& Detisch(2013)]{pfsdc} 
  R.~L.~Porter, G.~J.~Ferland, P.~J.~Storey and M.~J.~Detisch,
  Mon.\ Not.\ Roy.\ Astron.\ Soc.\  {\bf 433}, L89 (2013)
  arXiv:1303.5115 [astro-ph.CO].
  
\bibitem[Benjamin, Skillman, \& Smits(2002)]{bss02}
  R.~A.~Benjamin, E.~D.~Skillman and D.~P.~Smits,
   Astrophys.\ J.\  {\bf 569}, 288 (2002)
  [astro-ph/0202227].

\bibitem[Ferland et al.(2013)]{fer13}
  G.~J.~Ferland, R.~L.~Porter, P.~A.~M.~van Hoof,  R.~J.~R.~Williams, N.~P.~Abel, M.~L.~Lykins, G.~Shaw, W.~J.~Henney and P.~C.~Stancil,
  RevMexAA {\bf 49}, 137 (2013)
  arXiv:1302.4485 [astro-ph.GA].

\bibitem[Porter et al.(2009)]{pfms09} 
  R.~L.~Porter, G.~J.~Ferland, K.~B.~MacAdam and P.~J.~Storey,
  Mon.\ Not.\ Roy.\ Astron.\ Soc.\  {\bf 393}, L36 (2009)
  arXiv:0811.1216 [astro-ph].

\bibitem[Hummer \& Storey(1998)]{hs98} 
  D.~G.~Hummer and P.~J.~Storey, 
  Mon.\ Not.\ Roy.\ Astron.\ Soc.\   {\bf 297}, 1073 (1998)

\bibitem[Vriens \& Smeets(1980)]{vs80} 
  L.~Vriens and A.~H.~M.~Smeets,
  Phys.\ Rev.\ A {\bf 22}, 940 (1980).


\bibitem[Izotov et al.(2013)]{isg13} 
  Y.~I.~Izotov, G.~Stasinska and N.~G.~Guseva,
  arXiv:1308.2100 [astro-ph.CO].

\bibitem[Skillman et al.(2013)]{leop} 
  E.~D.~Skillman, J.~J.~Salzer, D.~A.~Berg, R.~W.~Pogge, N.~C.~Haurberg, J.~M.~Cannon, E.~Aver and K.~A.~Olive {\it et al.},
  Astrophys.\ J.\  {\bf 146}, 16 (2013)
  arXiv:1305.0277 [astro-ph.CO].


\end{thebibliography}
\end{document}